\documentclass[aps,twocolumn,amsmath,letterpaper]{revtex4-1}
\usepackage{amssymb}
\usepackage{graphicx}
\usepackage{array}
\usepackage{hhline}
\usepackage{longtable}

\begin{document}

    \title{Deep Spin-Glass Hysteresis Area Collapse and Scaling in the $d=3$ $\pm J$ Ising Model}

    \author{Ozan S. Sar{\i}yer$^{1}$, Alkan Kabak\c{c}{\i}o\u{g}lu$^{1}$, and A. Nihat Berker$^{2,3}$}
    \affiliation{$^1$Department of Physics, Ko\c{c} University, Sar{\i}yer 34450, Istanbul, Turkey,}
    \affiliation{$^2$Faculty of Engineering and Natural Sciences, Sabanc\i~University, Orhanl\i , Tuzla 34956, Istanbul, Turkey,}
    \affiliation{$^3$Department of Physics, Massachusetts Institute of Technology, Cambridge, Massachusetts 02139, U.S.A.}

\begin{abstract}
We investigate the dissipative loss in the $\pm J$ Ising spin glass
in three dimensions through the scaling of the hysteresis area, for
a maximum magnetic field that is equal to the saturation field. We
perform a systematic analysis for the whole range of the bond
randomness as a function of the sweep rate, by means of
frustration-preserving hard-spin mean field theory. Data collapse
within the entirety of the spin-glass phase driven adiabatically
(\emph{i.e.}, infinitely-slow field variation) is found, revealing a
power-law scaling of the hysteresis area as a function of the
antiferromagnetic bond fraction and the temperature. Two dynamic
regimes separated by a threshold frequency $\omega_c$ characterize
the dependence on the sweep rate of the oscillating field. For
$\omega < \omega_c$, the hysteresis area is equal to its value in
the adiabatic limit $\omega=0$, while for $\omega>\omega_c$ it
increases with the frequency through another randomness-dependent
power law.


\end{abstract}

\maketitle
\def\s{\rule{0in}{0.28in}}
\setlength{\LTcapwidth}{\columnwidth}

%\section{Introduction}
Hysteresis in magnetic materials has been a subject of interest for
quite some time due to applications in magnetic memory devices and
as a testground for theories of nonequilibrium
phenomena~\cite{Bertotti,Colaiori08,Pierce07,Katzgraber06}. The hysteresis area
which measures the magnetic energy loss in the material is connected
with the Barkhausen noise~\cite{Barkhausen,Durin05} due to
irreversible avalanche
dynamics~\cite{BTW87,Cote91,Levy93,Sethna93,Perkovic95,Hardy04}.
Existing literature on hysteresis in random magnets focuses mostly
on random-field
models~\cite{Perkovic95,Sabhapandit00,Dahmen01,Sethna05} while
numerical studies on random-bond models are mostly at zero
temperature~\cite{Vives94,Vives95,Vives95F,Hartmann99,Vives00,Pazmandi00,Katzgraber02}.
To our knowledge, there has been no finite-temperature study of the
hysteresis loss, especially in the spin-glass phase where large
avalanches are expected to be severely prohibited. We here
investigate the adiabatic and dynamic hysteresis in the the $\pm J$
random-bond Ising spin glass~\cite{Toulouse} on a finite,
3-dimensional simple cubic lattice with periodic boundary
conditions. We show that the hysteresis area obeys a scaling
relation in the whole spin-glass phase, in accord with earlier
theoretical studies which observe scale invariance over the whole
range about the critical disorder for various disorder-driven
systems~\cite{Sethna05,Vives94,Vives95}. Moreover, this scaling data
collapse is also observed for experimental systems over wide ranges
of temparature/magnetic-field scaling variables: Gingras \emph{et
al.} observed a universal data collapse for a
temparature/magnetic-field scaling variable over four decades in a
geometrically frustrated antiferromegnet
Y$_2$Mo$_2$O$_7$~\cite{Gingras97}, while Gunnarsson \emph{et al.}
observed such a data collapse for the short-range Ising spin glass
Fe$_{0.5}$Mn$_{0.5}$TiO$_3$~\cite{Gunnarsson91}.

The $\pm J$ Ising spin-glass model is defined by the dimensionless
Hamiltonian
\begin{equation}
  \label{eq:Ham}
  -\beta\mathcal{H}=\sum_{\langle ij\rangle}J_{ij}s_is_j+H\sum_is_i\:,
\end{equation}
where $\beta\equiv\frac{1}{k_BT}$ is the inverse temperature. The
first sum in Eq.(\ref{eq:Ham}) is over the pairs of nearest-neighbor
sites $(i,j)$, where $J_{ij}$ is the quenched-random local
interaction between the classical Ising spins $s_i = \pm1$.
The probability distribution function for $J_{ij}$ is given by
\begin{equation}
  \label{eq:distro}
  P(J_{ij})=p\,\delta(J_{ij}+J)+(1-p)\delta(J_{ij}-J)\:.
\end{equation}
$H$ in the second term in Eq.(\ref{eq:Ham}) is the uniform external
magnetic field. With a proper choice of units, the temperature for
the system may be defined as $T\equiv1/J$. A random distribution of
ferromagnetic and antiferromagnetic bonds gives rise to frustration
and yields a spin-glass phase for a range of $p$ values. Ising
spin-glass models are widely used as a tool for understanding
properties of experimental spin glasses such as
Pr$_{0.6}$Ca$_{0.4}$Mn$_{0.96}$Ga$_{0.04}$O$_{3}$~\cite{Hardy04},
Fe$_{0.5}$Mn$_{0.5}$TiO$_3$~\cite{Gunnarsson91,Katzgraber07,Ito86},
LiHo$_{0.167}$Y$_{0.833}$F$_4$~\cite{WenhaoWu93}, and
Cu$_{3-x}$AlMn$_x$~\cite{Vives00b}. Without loss of generality we
set $p \leq 0.5$, since the partition function is invariant under
the transformation $p,\{s_i^A\},\{s_j^B\} \rightarrow
(1-p),\{s_i^A\},\{-s_j^B\}$, where $A$ and $B$ signify the two
sublattices.

For small values of $p$ and $H=0$, the orientational (up-down)
symmetry is spontaneously broken below a critical temperature
$T_c(p)$ and long-range ferromagnetic order sets in. This phase is
well understood within the Landau picture where the free energy
landscape is described by two minima at magnetizations $\pm m(T,p)$.
Beyond a critical fraction $p_c$ of antiferromagnetic bonds,
reducing temperature drives the system into a glassy phase. The
low-temperature phase now retains its orientational symmetry and a
new, randomness-dominated phase which has a broken replica symmetry
appears~\cite{Parisi83,Mezard84}. In this phase, the free energy
landscape is rough, with many local minima at significantly
nonoverlapping configurations. Meanwhile, the dynamics slows down to
the extent that the relaxation time diverges~\cite{Binder86}. At
high temperatures, $T>T_c(p)$, both ordered phases give way to a
paramagnetic state where the entropic contribution to the free
energy is dominant. While the critical temperature strongly depends
on $p$ along the ferromagnet-to-paramagnet phase boundary, only a
weak dependence of $T_c$ on $p$ is observed for the spin-glass
phase~\cite{Binder86,Ozeki87}. In this study, we investigate the
hysteretic behavior of a spin glass under uniform magnetic field $H$
that is swept at a constant rate $\omega$. A past computational
study similar to ours~\cite{Yucesoy} considers a time-dependent
quenched-random magnetic field that is conjugate to the spin-glass
order parameter.

\begin{figure}[t!]
  \centering \includegraphics*[scale=1.0]{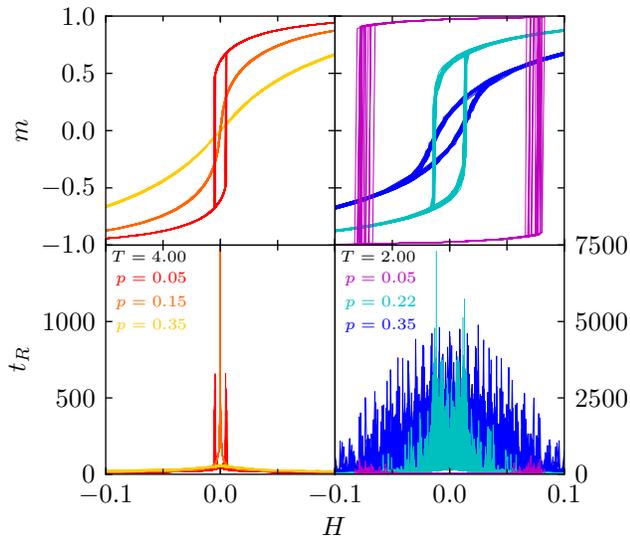}
  \caption{(Color) Hysteresis curves (upper) and relaxation times (lower) at
    high ($T\!=\!4.00$, left) and low ($T\!=\!2.00$, right)  temperatures.
    Data are for $p$ values either deep in the
    ferromagnetic ($p\!=\!0.05$), spin-glass ($T\!=\!2.00$, $p\!=\!0.35$) or paramagnetic ($T\!=\!4.00$, $p\!=\!0.35$)
    phases, or close to the phase boundaries for the ferromagnetic/paramagnetic
    ($T\!=\!4.00$, $p\!=\!0.15$) or ferromagnetic/spin-glass
    ($T\!=\!2.00$, $p\!=\!0.22$) transitions. For each case, an overlay of 20 distinct
    runs with different random-bond arrangements is shown.}
  \label{fig:hysteresis}
\end{figure}

We use hard-spin mean-field theory (HSMFT), a self-consistent field
theoretical
approach~\cite{Yucesoy,HSMFT01,HSMFT02,HSMFT03,HSMFT04,HSMFT05,HSMFT06,HSMFT07,HSMFT08,HSMFT09,HSMFT10,HSMFT11,HSMFT12,HSMFT13,HSMFT14,HSMFT15,HSMFT16}
that preserves the effects due to frustration (crucial for the
spin-glass phase) generated by the randomly scattered
antiferromagnetic bonds. HSMFT is defined by the refined set of
self-consistent equations
\begin{equation}
  \label{eq:hsmft}
  m_i=\sum_{\{s_j\}}\Bigg\{\Bigg[\prod_jP(m_j,s_j)\Bigg]\tanh\Bigg(\sum_jJ_{ij}s_j+H\Bigg)\Bigg\}
\end{equation}
for the local magnetization $m_i$ at each site $i$, whose
nearest-neighbors are labeled by $j$.  The single-site probability
distribution is
\begin{equation}
   P(m_j,s_j)=\frac{1+m_js_j}{2}\:.  \end{equation}
The local magnetization $m_i$ at site $i$ satisfies $-1\leq
m_i\leq1$. The hard-spin mean-field theory Eq.(3) has been discussed
in detail in
Refs.~\cite{Yucesoy,HSMFT01,HSMFT02,HSMFT03,HSMFT04,HSMFT05,HSMFT06,HSMFT07,HSMFT08,HSMFT09,HSMFT10,HSMFT11,HSMFT12,HSMFT13,HSMFT14,HSMFT15,HSMFT16}.

HSMFT has been successfully applied to spin
glasses~\cite{Yucesoy,HSMFT05}. In this paper we make use of the
method to investigate the scaling of the hysteresis area under a
uniform, time-dependent magnetic field. To this end, we consider a
$20\times 20\times 20$ cubic lattice with periodic boundary
conditions. We have checked in this study and in a previous
study~\cite{Yucesoy} that our hard-spin mean-field theory results
are independent of size for an $L\times L\times L$ system for
$L\gtrsim15$. A particular realization at a given $(T,p)$ is
generated by the assignment of the quenched-random coupling
constants $J_{ij}$ according to the probability distribution of
Eq.(\ref{eq:distro}) and, initially, a random and unbiased choice of
spins $s_i=\pm1$. In order to determine the hysteresis curves, the
system is first saturated by a sufficiently large external field
$H_s$, the minimum value of $H$ for which Eq.(\ref{eq:hsmft}) yields
an average magnetization $m = (1/L^3)\sum_i m_i = 1$ within an
accuracy $\epsilon_m \equiv 10^{-6}$. Then, the path $H_s\rightarrow
-H_s \rightarrow H_s$ is traversed with steps $\Delta H = H_s/100$ %% OZAN: onaylar misin? --> Dogrudur.
or smaller. For each incremental change of the field, the system is
allowed to relax a number of time steps $\tau = 1/\omega$. A time
step corresponds to successive iterations of Eq.(\ref{eq:hsmft}) on
$L^3$ arbitrarily chosen sites. An infinitely slow sweep is obtained
as the limit $\tau\to t_R$, where the HSMF equations converge to a
self-consistent solution within the tolerance interval $\epsilon_m$.
Thus, $t_R$ is the relaxation time of the system.

\begin{figure}[b!]
  \centering
  \includegraphics*[scale=1.0]{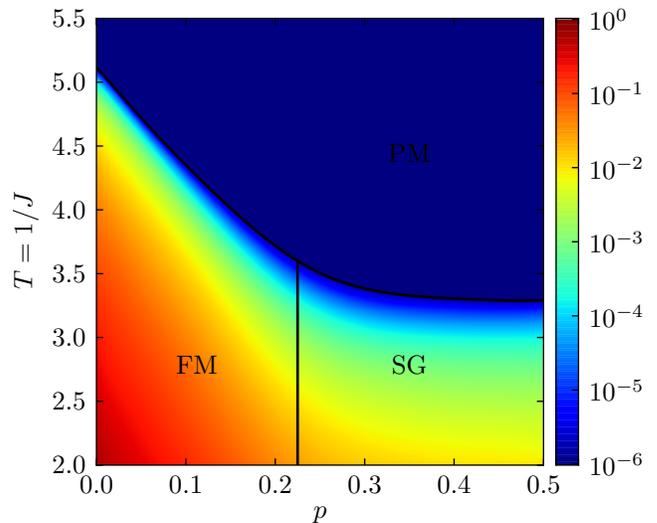}
  \caption{(Color) Logarithmic contour plot of the infinitely-slow-sweep hysteresis area $A_0$
    as a function of antiferromagnetic bond probability $p$ and temperature $T=1/J$.
    The thick vertical line denotes the phase boundary between the ferromagnetic and the
    spin-glass phases as described in the text, while the other thick line bounds the
    paramagnetic phase where the infinitely-slow-sweep hysteresis area is less than the
    precision used in the consistent-field calculations, \emph{i.e.}, $A_0<10^{-6}$.}
  \label{fig:contour}
\end{figure}

\begin{figure*}[t!]
\begin{tabular}{lr}
  \includegraphics*[scale=1.0]{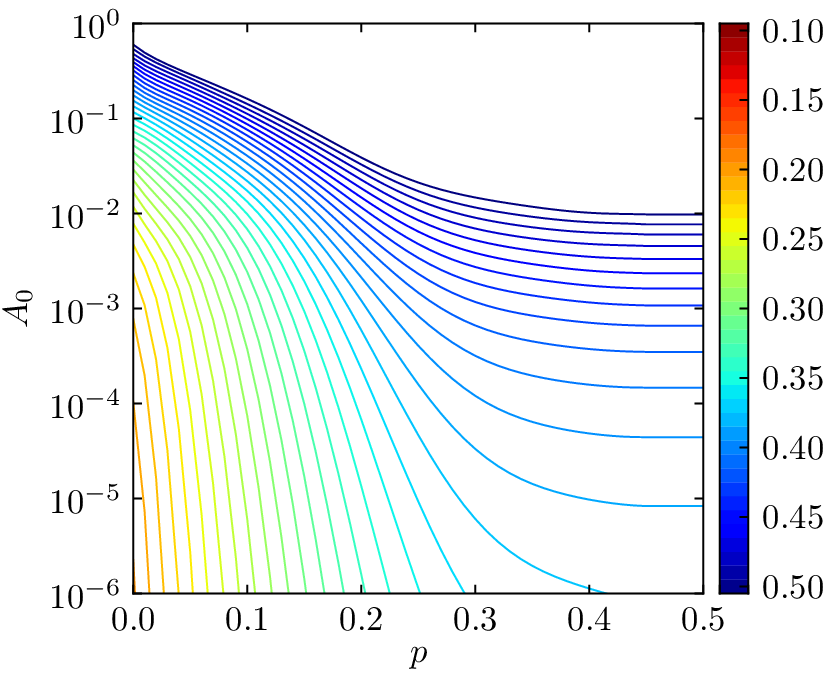} &  \includegraphics*[scale=1.0]{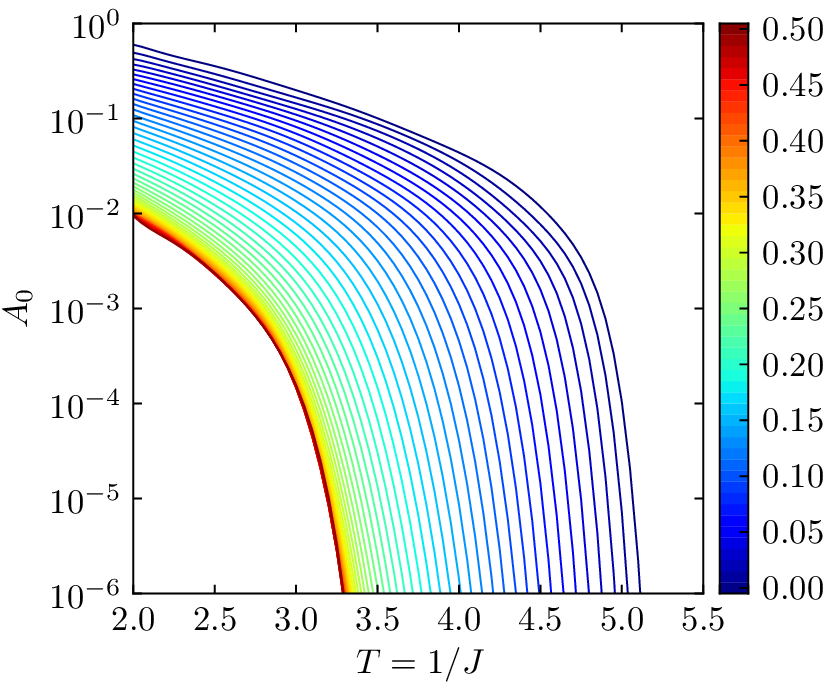}
\end{tabular}
 \caption{(Color) Infinitely-slow-sweep hysteresis area $A_0$, as a function
   of antiferromagnetic bond probability $p$ for temperatures
   (indicated in the color legend)
   $1/T=J=0.10$, $0.11$, $\ldots$, $0.50$ (left) and as a function of temperature $T=1/J$
   for antiferromagnetic bond probabilities (indicated in the color legend) $p=0.00$, $0.01$, $\ldots$, $0.50$ (right).
   Each curve is a 10th degree polynomial fit to the averages over 20 realizations.}
  \label{fig:JA} %  \label{fig:pA}
\end{figure*}

The infinitely-slow-sweep hysteresis curves obtained in the
ferromagnetic and spin-glass phases are shown in
Fig.\ref{fig:hysteresis}. The usual jump in the magnetization at a
coercive field $H_c$, observed for small $p$, is associated with a
system-wide avalanche in the ferromagnetic phase. For $p$ larger
than a critical value $p_c$, this picture is replaced by a slanted
hysteresis curve and a smaller hysteresis area, typical of
spin-glass materials~\cite{Pierce07,Hardy04,Vives00b}. This converse
hysteretic behavior, associated with the Barkhausen
noise~\cite{Barkhausen, Durin05}, is a consequence of the power-law
distribution of avalanches which is well
established~\cite{Durin05,BTW87,Sethna93,Hardy04,Perkovic95,Dahmen01,Sethna05,Vives94,Vives95,Vives95F,Vives00,Pazmandi00,Vives00b,Vives01}
for several frustrated systems with quenched disorder. The
hysteresis area disappears in the paramagnetic phase.

In Fig.\ref{fig:contour}, we present the infinitely-slow-sweep
hysteresis area globally, for all temperatures and antiferromagnetic
bond probabilities, on a logarithmic color-contour plot. The
hysteresis area $A_0$ vanishes in the region shown in dark blue,
which corresponds to the paramagnetic phase, while it is nonzero in
the ferromagnetic and spin-glass phases, respectively on the left
and right of the lower half of Fig.\ref{fig:contour}. The para/ferro
and para/spin-glass phase boundaries are easily determined by
locating the temperature at which $A_0$ vanishes (\emph{i.e.}, falls
below $\epsilon_m$). A set of $p$ scans for different temperatures
and a set of temperature scans for various $p$ values are given in
Fig.\ref{fig:JA}.  The low-temperature ferro/spin-glass boundary is
located at $p_c\simeq 0.22$ and is calculated as the inflection
point for the maximum slope of the hysteresis curve as a function of
antiferromagnetic bond probability~\cite{Vives94}. The phase
boundaries are consistent with the well-known phase diagram for the
3-dimensional $\pm J$ model~\cite{Ozeki87} and in fair comparison
with the experimental temperature-concentration phase diagrams of
the various Eu$_x$Sr$_{1-x}$S$_y$Se$_{1-y}$, solid
(o-H$_2$)$_{1-x}$(p-H$_2$)$_x$, and AuFe systems reviewed in
\cite{Binder86}.

\begin{figure}[b!]
  \centering
  \includegraphics*[scale=1.0]{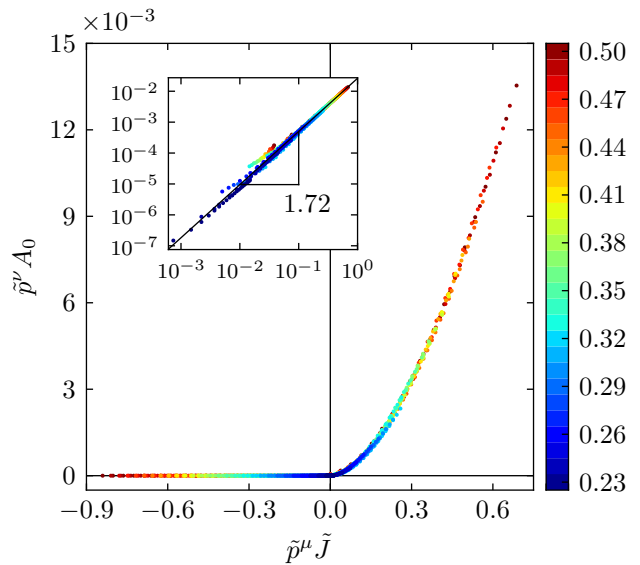}
  \caption{(Color) Scaling of the hysteresis area in the spin-glass phase as a
    function of reduced antiferromagnetic bond concentration $\tilde{p}$
    and the reduced bond strength $\tilde{J}$, for various $p$ values as shown
    in the color legend. The scaling function $f(x)$ given by the RHS of
    Eq.(\ref{eq:scaling}) on which all data points collapse is consistent
    with a same power law within the entire spin-glass phase.}
  \label{fig:scaling}
\end{figure}

We here focus on the scaling form of the hysteresis area in the
spin-glass phase and show that a unique scaling-function governs the
whole range of $p$ and $J$ within the spin-glass phase. To this end,
we first express the hysteresis area in the form
$A_0=A_0(\tilde{p},\tilde{J})$, where
$\tilde{p}\equiv\frac{p-p_c}{p_c}$ and
$\tilde{J}\equiv\frac{J-J_c}{J_c}$ are the reduced displacements from
phase boundaries. We then postulate the multivariate scaling form
\begin{equation}
  A_0(\tilde{p},\tilde{J})=\lambda^cA_0(\lambda^a\tilde{p},\lambda^b\tilde{J})\:,
\end{equation}
which by letting $\lambda=\tilde{p}^{-1/a}$ reduces to
\begin{equation}
  A_0(\tilde{p},\tilde{J})=\tilde{p}^{-c/a}A_0(1,\tilde{p}^{-b/a}\tilde{J})\:,
\end{equation}
Defining $\nu\equiv c/a$, $\mu\equiv-b/a$, and $f(x)\equiv
A_0(1,x)$, we obtain
\begin{equation}
  \tilde{p}^\nu A_0(\tilde{p},\tilde{J})=f(\tilde{p}^\mu\tilde{J})\:.
  \label{eq:scaling}
\end{equation}

The sought collapse is obtained by the choice of scaling exponents
$\mu=1$ and $\nu=2$.  Data shown in Fig.\ref{fig:JA} collapses onto
a single curve shown in Fig.\ref{fig:scaling}, where the LHS of
Eq.(\ref{eq:scaling}) is plotted against the argument on the RHS for
28 evenly spaced values of $p$ above $p_c$. The origin corresponds
to the phase boundary between the spin-glass and paramagnetic
phases. The log-log plot of the same collapse shown in the inset of
Fig.\ref{fig:scaling} suggests that the scaling function has the
form $f(x) \propto x^{1.72}$, yielding a hysteresis area $A_0
\propto \tilde{p}^\alpha\tilde{J}^\beta$ with $\alpha\simeq -0.28$
and $\beta\simeq 1.72$. Interestingly, unlike the case of usual
critical phenomena, the scale-invariance applies to the entire
spin-glass phase and not just to the vicinity of the critical phase
boundary.

\begin{figure}%[t!]
  \centering
  \includegraphics*[scale=1.0]{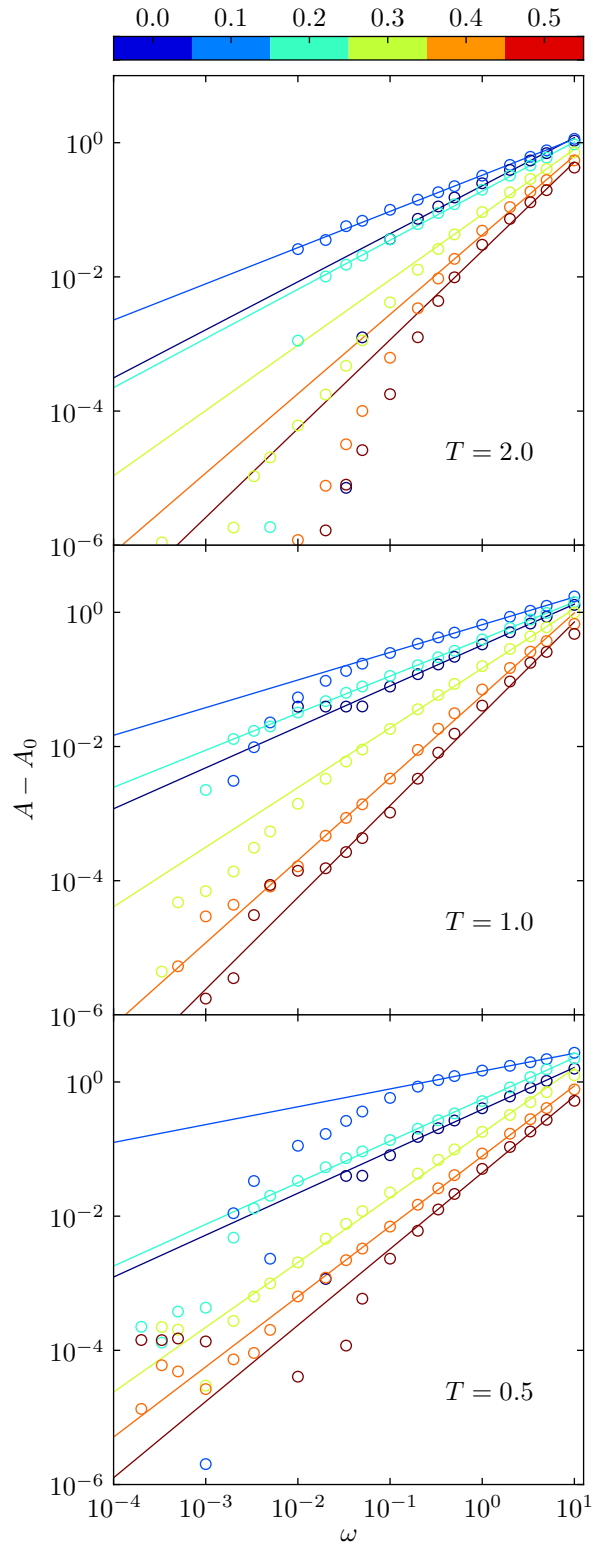}
  \caption{(Color) Hysteresis area difference $A-A_0$ versus sweep rate $\omega$, for
  temperatures $T=2.0$, $1.0$, $0.5$ from top to bottom and for antiferromagnetic
  bond fractions $p=0.0$, $0.1$, $\ldots$, $0.5$ as shown in the color legend.}
  \label{fig:sweep_rate_area}
\end{figure}

Having analyzed the limit with infinitely-slow-sweep rate, we next
consider the dynamic hysteretic response as a function of the
magnetic field frequency. One can simulate the finite oscillation
frequency by iterating Eq.(\ref{eq:hsmft}) for a predetermined
number of steps $t$, instead of waiting until a steady-state is
reached. The sweep rate $\omega = 1/t$ is proportional to the
frequency of the applied field up to a material-dependent spin
relaxation time. The hysteresis area $A(\omega,p,J)$ deviates from
the value at infinitely-slow sweep $A_0=A(\omega\!\!=\!\!0, p, J)$
and increases with increasing sweep rate $\omega$. This can be
understood by observing that the laggy response of the magnetization
to a time-varying field inflates the hysteresis curve along the
field direction. The typical behavior observed in various
experimental and theoretical magnets (typically pure magnets or
random-field systems)~\cite{Jung, Zheng1, Zheng2, Zheng3, Acharyya}
is
\begin{equation}
  \label{eq:SweepRateExp}
  A(\omega,p,J)=A_0+g(p,J)\: \omega^b\:,
\end{equation}
where $b$ is the sweep-rate exponent. We investigate whether the
random-bond Ising spin glass obeys a similar scaling relation.

\begin{figure}%[b!]
  \centering
  \includegraphics*[scale=1.0]{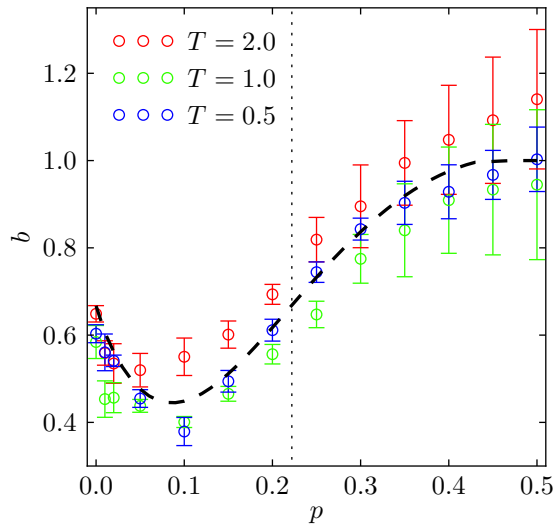}
  \caption{(Color) Sweep-rate exponent $b$ versus antiferromagnetic
  bond fraction $p$ for temperatures $T=2.0$, $1.0$, and $0.5$.
  The dashed curve depicts the general trend of the sweep-rate
  exponent, while the dotted vertical line marks the phase transition
  from ferromagnetic to spin-glass phase.}
  \label{fig:sweep_rate_exp}
\end{figure}

A typical scan of the hysteresis area as a function of $\omega$
displays two dynamic regimes, separated by a critical sweep rate
$\omega_c$ that depends on $p$, $J$, and the system size
(Fig.\ref{fig:sweep_rate_area}). For a sufficiently slowly varying
field $\omega<\omega_c$, the area is pinned at the value $A_0$. In
this regime, the avalanches that are triggered by an incremental
increase in the field decay within a period $1/\omega$ or smaller.
For faster sweeps ($\omega>\omega_c$), the increase in the area
follows the power law in Eq.(\ref{eq:SweepRateExp}), with a
$p$-dependent exponent $b$. In the ferromagnetic phase with weak
disorder, the two dynamic regimes are separated by a sharp increase
in the hysteresis area. This transition gets significantly smoother
in the spin-glass phase, especially far from the
ferromagnetic/spin-glass boundary. For larger systems, one expects
$\omega_c$ to recede and the power-law behavior to dominate.

Fig.\ref{fig:sweep_rate_exp} shows the sweep-rate exponent $b$
calculated as a function of antiferromagnetic bond fraction $p$, at
fixed temperatures $T=1/J=2.0$, $1.0$, and $0.5$. The hysteresis area is calculated for
the sweep rates $\omega = 1$, $0.5$, $0.\bar{3}$, $0.2$, $10^{-1}$,
$\ldots$, $10^{-4}$ at each $p$ value, after averaging over 10
realizations. The exponent values are obtained through fits to the
data in the regime $\omega>\omega_c$ (typically two decades or
more), using the functional form of Eq.(\ref{eq:SweepRateExp}). The
error bars reflect only the scatter of the data relative to the fit.
In the ferromagnetic phase, $p<p_c$, we note that the calculated
sweep-rate exponents lie in an interval of fairly good agreement
with the various values obtained previously at $p=0$, namely
$b=2/3$~\cite{Jung, Zheng1, Zheng2, Zheng3} and
$b=0.52\pm0.04$~\cite{Zheng1} from mean-field theory,
$b=0.61$~\cite{Zheng1} from Glauber dynamics simulations,
$b=0.495\pm0.005$~\cite{Zheng2} and $b=0.45$~\cite{Acharyya} from
Monte Carlo simulations.

In conclusion, we have considered here the $\pm J$ Ising model under
uniform external field and investigated the scaling behavior of the
saturation hysteresis area (\emph{i.e.}, far from the weak-field
limit). We observed that the phase diagram can be derived from the
hysteresis area alone and the ferromagnetic/spin-glass phase
boundary corresponds to the inflection point {\it wrt}
bond-randomness strength $p$. When adiabatically driven, the area
displays a data collapse within the entire spin-glass phase for all
temperatures and $p$. The scaling function itself has a power-law
form and the scale-invariance extends far from the phase boundary,
deep into the spin-glass phase.

The dynamical response under a fluctuating external field is also
interesting. We find that, beyond a threshold value $\omega_c$, the
hysteresis area increases as a function of the field-sweep rate
$\omega$  with a non-universal power-law. This behavior is not
limited to the vicinity of the phase transition. The associated
exponent is found to be a function the randomness strength $p$.
Moreover, this function is independent of temperature. In the limit
of a pure magnet ($p\to 0$), we observe good agreement with the
existing literature, despite the fact that the earlier theoretical
work applies to a weak driving field, while we here consider sweeps
across saturation limits. Fig.\ref{fig:sweep_rate_exp} suggests
that, relative to the ferromagnetic phase, the spin glass displays
an amplified sensitivity to the field-sweep rate, again running in
apparent contrast with the general wisdom that the hysteretic
effects are suppressed within a spin glass. In fact, we note that
the increase in the hysteresis area with $\omega$ is due to the
magnet's delayed response to the changing field, and a signature of
the spin-glass phase is the slowing down of precisely such relaxation phenomena.\\

\begin{acknowledgments}
Support by the Alexander von Humboldt Foundation, the Scientific and
Technological Research Council of Turkey (T\"UB\.ITAK), and the
Academy of Sciences of Turkey (T\"UBA) is gratefully acknowledged.
We acknowledge the hospitality of the T\"UB\.ITAK-Bosphorus
University Feza G\"{u}rsey Institute for Fundemental Sciences, for
the computational support from the Gilgamesh cluster.
\end{acknowledgments}

\end{document}